\documentclass{nature}
\title{How Natural Selection Can Create Both Self- and Other-Regarding Preferences, and Networked Minds}

\author{Thomas Grund$^{1,2}$, Christian Waloszek$^{1}$ and Dirk Helbing$^{1\ast}$}

\begin{document}

\maketitle

%%%%%%%%%%%%%%%%%%%%%%%%%%%%%%%%%%%%%%%%%%%%%%%%%%%%%%%%%%%%%%%%
%%%%%%%%%%             AFFILIATIONS             %%%%%%%%%%%%%%%%%%%%%%%%%%%%%%%%
%%%%%%%%%%%%%%%%%%%%%%%%%%%%%%%%%%%%%%%%%%%%%%%%%%%%%%%%%%%%%%%%

\begin{affiliations}
 \item Chair of Sociology, in particular of Modeling and Simulation, ETH Zurich, Switzerland
 \item Centre International de Criminologie Compar\'{e}e, Universit\'{e} de Montr\'{e}al, Canada
\end{affiliations}

%%%%%%%%%%%%%%%%%%%%%%%%%%%%%%%%%%%%%%%%%%%%%%%%%%%%%%%%%%%%%%%%
%%%%%%%%%%            ABSTRACT            %%%%%%%%%%%%%%%%%%%%%%%%%%%%%%%%%%%
%%%%%%%%%%%%%%%%%%%%%%%%%%%%%%%%%%%%%%%%%%%%%%%%%%%%%%%%%%%%%%%%

\begin{abstract}
Biological competition is widely believed to result in the evolution of selfish preferences. The related concept of the `homo economicus' is at the core of mainstream economics. However, there is also experimental and empirical evidence for other-regarding preferences. Here we present a theory that explains both, self-regarding and other-regarding preferences. Assuming conditions promoting non-cooperative behaviour, we demonstrate that intergenerational migration determines whether evolutionary competition results in a `homo economicus' (showing self-regarding preferences) or a `homo socialis' (having other-regarding preferences). Our model assumes spatially interacting agents playing prisoner's dilemmas, who inherit a trait determining `friendliness', but mutations tend to undermine it. Reproduction is ruled by fitness-based selection without a cultural modification of reproduction rates. Our model calls for a complementary economic theory for `networked minds' (the `homo socialis') and lays the foundations for an evolutionarily grounded theory of other-regarding agents, explaining individually different utility functions as well as conditional cooperation. 
\end{abstract}
\clearpage

%%%%%%%%%%%%%%%%%%%%%%%%%%%%%%%%%%%%%%%%%%%%%%%%%%%%%%%%%%%%%%%%
%%%%%%%%%%               KEYWORDS               %%%%%%%%%%%%%%%%%%%%%%%%%%%%%%%%
%%%%%%%%%%%%%%%%%%%%%%%%%%%%%%%%%%%%%%%%%%%%%%%%%%%%%%%%%%%%%%%%

%% \keywords{other-regarding preferences | evolution | conditional cooperation | fairness | homo economicus | homo socialis} # not used in Nature Scientific Reports

%%%%%%%%%%%%%%%%%%%%%%%%%%%%%%%%%%%%%%%%%%%%%%%%%%%%%%%%%%%%%%%%
%%%%%%%%%%            TEXT of ARTICLE            %%%%%%%%%%%%%%%%%%%%%%%%%%%%%%%
%%%%%%%%%%%%%%%%%%%%%%%%%%%%%%%%%%%%%%%%%%%%%%%%%%%%%%%%%%%%%%%%

Many societal problems, such as pollution, global warming, overfishing, or tax evasion, result from social dilemmas. In these dilemmas, uniform cooperation would be good for everybody, but each individual can benefit from free-riding\cite{1}. Although societies found ways to cope with so-called `tragedies of the commons'\cite{2}, the evolution of other-regarding preferences under competitive selection pressure is still a challenging and topical scientific puzzle.

In social dilemma situations, caring about others can reduce individual success. While profit maximisation in single interactions would always demand non-cooperative behaviour, repeated interactions may sometimes support reciprocal altruism and result in human sociality\cite{3,4}. But even in one-shot interactions, humans are not as selfish as theory suggests. A large body of experimental and field evidence indicates that people genuinely care about each other\cite{5,6,7,8}. They tend to be not only concerned about individual success, but also about that of others\cite{6,8,9}. 

Also in dictator and ultimatum games, the tendency to share is often attributed to other-regarding preferences\cite{10,other}. But how did such other-regarding preferences evolve and spread? It was suggested that group selection would solve the puzzle\cite{11,12,add2} but it only works when groups do not mix\cite{13}. Therefore, mechanisms not requiring kin or group selection have been looked for\cite{extra}. Some authors argue that humans intrinsically favour fairness\cite{14,10}.  But why do other-regarding preferences then vary geographically\cite{6}?

To explain an innate sense of fairness, Gintis proposed a modification of reproduction rates by culture\cite{15}. Other models studying the evolution of fairness preferences typically assume mechanisms in favour of pro-sociality, such as a `shadow of the future'\cite{4}, costly punishment\cite{16,add4}, reputation\cite{17,18}, genetic favouritism\cite{19,20}, genetic drift\cite{21}, or local interactions with an imitation of more successful neighbours\cite{22,23,24,25,add3}. However, the `best response' rule is not favourable for the spreading of cooperation in social dilemma situations, where non-cooperative behaviour creates a higher payoff, no matter what the behavioural strategy of the interaction partner(s) is\cite{26}.

%%%%%%%%%%%%%%%%%%%%%%%%%%%%%%%%%%%%%%%%%%%%%%%%%%%%%%%%%%%%%%%%
%%%%%%%%%%               MODEL               %%%%%%%%%%%%%%%%%%%%%%%%%%%%%%%%%%
%%%%%%%%%%%%%%%%%%%%%%%%%%%%%%%%%%%%%%%%%%%%%%%%%%%%%%%%%%%%%%%%

\section*{Results}

\subsection{Model.}

Our model does not require any of the previously mentioned social mechanisms and it even works when the best response rule is applied. For simplicity, we assume a spatial square lattice with periodic boundary conditions and $L\times L$ sites; 60\% of the sites are occupied by one agent each, the other 40\% are empty. Agents simultaneously interact with all other agents in their Moore neighbourhood---the eight sites surrounding their own site. In each time period, agents can choose to cooperate ($C$) or to defect ($D$). For all interactions with neighbours, agents get a payoff. If two interacting agents cooperate, each obtains the amount $R$ (`Reward'); if both defect, each gets $P$ (`Punishment'); and if one cooperates and the other one defects, the former one gets $S$ (`Sucker's Payoff'), while the latter gets $T$ (`Temptation'). The agents' reproductive fitness in time period t is given by the sum of all payoffs from interactions with neighbours (minus a value of $8|S|$ to ensure non-negative payoffs and avoid the reproduction of agents who are exploited by all their neighbours). 

At the end of each period, individuals die with probability $\beta$. To keep population size constant, all agents who die are replaced by an offspring of one of the surviving agents. The likelihood of parents to create an offspring is strictly proportional to their actual payoff, i.e. their reproductive fitness. The offspring is born in one of the empty sites closest to the parent with probability $\nu$ (`local reproduction'), while it occupies a random empty site irrespective of the distance to the parent with probability $1-\nu$ (`random reproduction').

We assume a strict prisoner's dilemma with $T > R > P > S$. Although collective success is highest when everybody cooperates, defection is the payoff-maximising individual strategy in each single interaction, independently of the neighbours' strategies. In our model, individuals update their strategy (cooperation or defection) based on the myopic best response rule at the end of each period. However, rather than maximising their payoff $P_i$, we assume here that an individual $i$ chooses the strategy that maximises the utility 
\begin{equation}
U_i =  (1-\rho_i)P_i + \rho_i \overline{P}, 
\end{equation}
where $\overline{P}$ denotes the average payoff of the interaction partners $j$. We do this because of previous studies and empirical evidence\cite{27,10,25} suggesting that the utility is not just given by the own payoff $P_i$, but the payoff $P_j$ of interaction partners $j$ is also given a certain weight $\rho_i$. The variable $\rho_i\in[0,1]$---the `friendliness'---represents the degree of other-regarding preferences of agent $i$. A purely self-regarding individual with $\rho_i=0$ only cares about the own payoff when choosing a strategy. An other-regarding individual gives the own payoff a weight of $1-\rho_i$ and the payoff of interaction partners a weight of $\rho_i$. Hence, strategy updates are assumed to be `empathic', but reproduction is exclusively driven by individual payoff. 

When selfishness is fixed ($\rho_i=0$), best response behaviour promotes a `tragedy of the commons'\cite{1}. Instead, however, we assume a (genetic or cultural) transmission of friendliness $\rho_i$ from parent $i$ to offspring $j$, which is subject to random mutation. In our model, mutation occurs with a constant probability $\mu$ that is independent of the strategies pursued in the neighbourhood. To avoid `genetic drift', which would eventually promote friendliness scores of 0.5, the mutation of $\rho_i$ is specified such that the offspring tends to be more self-regarding than the parent (if $\rho_i > 0.2$): With probability 0.8, $\rho_j$ is set to a uniformly distributed random value between 0 and $\rho_i$, and with probability 0.2 it is set to a uniformly distributed value between $\rho_i$ and 1.

%%%%%%%%%%%%%%%%%%%%%%%%%%%%%%%%%%%%%%%%%%%%%%%%%%%%%%%%%%%%%%%%
%%%%%%%%%%            RESULTS & DISCUSSION            %%%%%%%%%%%%%%%%%%%%%%%%%%
%%%%%%%%%%%%%%%%%%%%%%%%%%%%%%%%%%%%%%%%%%%%%%%%%%%%%%%%%%%%%%%%

\subsection{Simulation Results.}
Our computer simulations start in the most adverse condition for friendliness and cooperation. At time $t=0$, all agents defect and nobody cares about the payoff of others ($\rho_i=0$). However, mutations will eventually create higher levels of friendliness. According to the best response rule, an agent will cooperate, if the utility $U_i ( C ) $ of cooperation is larger than the utility $U_i(D)$ of defection. The utility of cooperation is 
\begin{equation}
U_i ( C ) = d[\rho_i T+(1-\rho_i)S]+cR, 
\end{equation}
when surrounded by $c$ co-operators and $d$ defectors, and the utility of defection is 
\begin{equation}
U_i(D)=dP+c[\rho_i S + (1-\rho_i)T]. 
\end{equation}
Therefore, cooperation is expected to occur for 
\begin{equation}
\frac{c}{d} > \frac{P-\rho_i T - (1-\rho_i)S}{R-\rho_i S - (1-\rho_i)T}. 
\end{equation}
That is, cooperativeness depends on the number of cooperative and defective neighbours, but it also depends on the level of friendliness  
$\rho_i$. We find that, for $\rho_i=0$, agents {\it never} cooperate, while above a critical threshold of friendliness, namely for 
\begin{equation}
\rho_i > \frac{P-S}{T-S}, 
\end{equation}
they cooperate unconditionally. For values $\rho_i$ of friendliness between $(T-R)/(T-S)$ and\newline $(P-S)/(T-S)$, we find {\it conditional} cooperation\cite{9}, when enough neighbours cooperated in the previous round (note that, in our simulations, $S<0$). 

Hence, `idealists' with a level of friendliness $\rho_i > (P-S)/(T-S)$ happen to cooperate even when they are surrounded and exploited by defectors. However, such idealists will normally get miserable payoffs and have very small reproduction rates. They tend to die without reproducing themselves. In fact, other-regarding preferences do not spread and selfishness thrives when offspring occupy randomly selected empty cells. 

In contrast, when agents reproduce locally, other-regarding preferences suddenly emerge after some time (see Fig. 1A). How does this surprising, sudden transition from the `homo economicus' to the `homo socialis' take place? In principle, mutations could create a random co-location of mutation-borne `idealists' by coincidence after a long time\cite{24}. 
This would lead to the formation of a cluster of cooperators of `supercritical' size. Such clustering would dramatically increase the relative fitness of other-regarding agents in the cluster and create sufficiently high reproduction rates to spread friendliness.

However, why does this transition happen in just a few generations (see Fig. 1B), i.e. much faster than expected? This relates to our distinction of preferences and behaviour. When an `idealist' is born in a neighbourhood with friendliness levels supporting conditional cooperation, this can trigger off a cascade of changes from defective to cooperative behaviour. Under such conditions, a single `idealist' may quickly turn a defective neighbourhood into a largely cooperative one. This implies higher payoffs and higher reproduction rates for both, idealists and conditional co-operators.

The intriguing phase transition from self-regarding to other-regarding preferences critically depends on the local reproduction rate (see Fig. 2). The clustering of friendly agents, which promotes other-regarding preferences, is not supported when offspring move away. Then, offspring are more likely to encounter defectors elsewhere and parents are not `shielded' by their own friendly offspring anymore. In contrast, with local reproduction, offspring settle nearby, and a clustering of friendly agents is reinforced. Under such conditions, friendliness is evolutionary advantageous over selfishness. 

\section*{Discussion}

In conclusion, we offer an over-arching theoretical perspective that could help to overcome the historical controversy in the behavioural sciences between largely incompatible views about human nature. Both, self-regarding and other-regarding types of humans may result from the same evolutionary process. Whereas high levels of intergenerational migration promote the evolution of a `homo economicus', low levels of intergenerational migration promote a `homo socialis', even under `Darwinian' conditions of a survival of the fittest and random mutations. The significance of local reproduction for the evolution of other-regarding preferences is striking and may explain why such preferences are more common in some parts of the world than in others\cite{6}.

Our modelling approach distinguishes between the evolution of individual preferences and behaviours. This makes cooperation conditional on the level of cooperation in the respective neighbourhood. Hence, when a few `idealists' are born, who cooperate unconditionally, this can trigger off cooperation cascades, which can largely accelerate the spreading of cooperation\cite{28}.  Our model can also serve as a basis to develop an economic theory of other-regarding agents. The advantage is that it does not need to assume certain properties of boundedly rational agents---these properties rather result from an evolutionary process. In fact, our model naturally explains the evolution of individually different utility functions, as they are experimentally observed (see Fig. 3 + 4), and also the evolution of conditional cooperators\cite{9, 29}.

A great share of economic literature is based on the assumption of the `homo economicus', who takes decisions without considering the payoff or utility of others. In contrast to this traditional view, the `homo socialis' never takes independent decisions, if the behaviour has external effects\cite{Christakis,Ockenfels}.
We might characterise this as a situation of `networked minds', where everybody is trying to put himself or herself into other people's shoes, to take into account their utilities in the decision-making process. As a consequence, besides paying attention to networks of companies\cite{Schweitzer}, economics should also consider networks of individual minds, i.e. social aspects. This is of particular relevance for information societies, in which individuals are increasingly connected via information and communication systems, such as social media\cite{FuturICT,Nature}. A theory of networked minds could make a significant contribution to the convergence of the behavioural sciences\cite{Gint}, and it might also shed new light on social capital, power, reputation and value, and create a fundamentally new understanding of these\cite{economics}. We believe that this view can stimulate a huge and exciting field of research, and lead to a complementary theory to the one based on the `homo economicus'. Due to the simplicity and fundamental character of the model proposed by us, we expect that it might serve as a starting point and basis for this new field.

%%%%%%%%%%%%%%%%%%%%%%%%%%%%%%%%%%%%%%%%%%%%%%%%%%%%%%%%%%%%%%%%
%%%%%%%%%%            BIBLIOGRAPHY            %%%%%%%%%%%%%%%%%%%%%%%%%%%%%%%%
%%%%%%%%%%%%%%%%%%%%%%%%%%%%%%%%%%%%%%%%%%%%%%%%%%%%%%%%%%%%%%%%

%%%%%%%%%%%%%%%%%%%%%%%%%%%%%%%%%%%%%%%%%%%%%%%%%%%%%%%%%%%%%%%%
%%%%%%%%%%            ADDENDUM            %%%%%%%%%%%%%%%%%%%%%%%%%%%%%%%%%%
%%%%%%%%%%%%%%%%%%%%%%%%%%%%%%%%%%%%%%%%%%%%%%%%%%%%%%%%%%%%%%%%
\begin{addendum}
 \item We would like to thank Thomas Chadefaux, Michael M\"as, and Ryan Murphy for valuable comments. This work was partially supported by the Future and Emerging Technologies program FP7-COSI-ICT of the European Commission through the project QLectives (Grant No. 231200) and by the ERC Advanced Investigator Grant `Momentum' (Grant No. 324247). DH is grateful to Stefan Rustler for preparing Fig. 3 and to Ryan Murphy for providing Fig. 4.
 \item [Author Contributions] DH and TG developed the model and wrote the paper. TG and CW performed the computer simulations and the data analysis.
 \item[Competing Interests] The authors declare no conflict of interests.
 \item[Correspondence] Correspondence and requests for materials
should be addressed to Dirk Helbing (email: dirk.helbing@gess.ethz.ch).
\end{addendum}

%%%%%%%%%%%%%%%%%%%%%%%%%%%%%%%%%%%%%%%%%%%%%%%%%%%%%%%%%%%%%%%%
%%%%%%%%%%                TABLES                %%%%%%%%%%%%%%%%%%%%%%%%%%%%%%%%%%
%%%%%%%%%%%%%%%%%%%%%%%%%%%%%%%%%%%%%%%%%%%%%%%%%%%%%%%%%%%%%%%%

%%%%%%%%%%%%%%%%%%%%%%%%%%%%%%%%%%%%%%%%%%%%%%%%%%%%%%%%%%%%%%%%
%%%%%%%%%%               FIGURES               %%%%%%%%%%%%%%%%%%%%%%%%%%%%%%%%%%
%%%%%%%%%%%%%%%%%%%%%%%%%%%%%%%%%%%%%%%%%%%%%%%%%%%%%%%%%%%%%%%%
\clearpage

\textbf{Figure 1}
A random spatial coincidence of friendly agents can lead to the sudden spreading of other-regarding preferences and a transition from a `homo economics' to a `homo socialis'. The graphs show representative simulation runs on a $30\times 30$ spatial grid with periodic boundary conditions. 60\% of all sites are occupied with agents who can either cooperate or defect. The payoff of interacting agents is determined as sum of payoffs from prisoner dilemma games with all Moore neighbours. The payoff parameters are: `Temptation' $T = 1.1$, `Reward' $R = 1$, `Punishment' $P = 0$, and `Sucker's Payoff' $S = -1$. The strategies (cooperation or defection) are updated simultaneously for all agents, applying the myopic best response rule to the utility function of each agent. It weights the payoffs of neighbours with the friendliness $\rho_i$ and the own payoff with $(1-\rho_i)$. Agents die at random with probability $\beta=0.05$. To keep population size constant, surviving agents produce offsprings proportionally to their payoff in the previous round. Offspring move to the closest empty site ($\nu=1$) and inherit attributes from the parent, here: the friendliness $\rho_i$. However, with probability $\mu=0.05$, the friendliness of offsprings mutates. With probability 0.8 it is `reset' to a uniformly distributed random value between 0 and the friendliness $\rho_i$ of the parent, and with probability 0.2 it takes on a uniformly distributed value between $\rho_i$ and 1. (A) Average of friendliness and share of cooperating agents as a function of time (one generation is $1/\beta$ periods). (B) Average payoffs of cooperators and defectors as a function of time. Initially, defectors are more successful than cooperators. However, when the sudden transition from a `homo economics' to a `homo socialis' occurs, the payoffs for defectors increases, but the payoffs for cooperators increases even more, which implies higher production rates of agents with other-regarding preferences.

\textbf{Figure 2}
Local reproduction is crucial for the transition from a `homo economicus' to a `homo socialis'. The rate $\nu$ of local reproduction determines the probability of an offspring to occupy the closest empty site to the parent. With probability $(1-\nu)$, the offspring moves to an empty site that is randomly selected. All other parameters are specified as in Fig. 1. The circle size indicates average friendliness, while the circle colour represents the share of cooperators. The values are averages over 100 simulation runs between generation 100 and 500. Even at temptation levels around $T = 1.3$, the above phase diagram shows a sudden transition from self-regarding preferences (small dots) to other-regarding preferences (large circles), when the degree $\nu$ of local reproduction is high enough.

\textbf{Figure 3}
Evolution of the distribution of friendliness in the course of time for the parameter values used in Fig. 1. The plot shows an average over 100 runs, smoothed with MATLAB's local regression using weighted linear least squares and a 1st degree polynomial model.
It is clearly visible that a broad distribution of individual utility functions results, even though everybody starts off with a purely self-regarding behaviour, for which the utility function agrees exactly with the payoff function.

\textbf{Figure 4}
Empirical distribution of other-regarding preferences (from R. Murphy et al. 2011, reproduction with kind permission of Ryan Murphy). \textit{(A)} This figure shows the primary items from a `Slider Measure' to determine Social Value Orientation. \textit{(B)} Distribution of Social Value Orientation scores from the experimentally determined Slider Measure, as represented by angles. The dark line is a smoothed kernel density estimation.

%%%%%%%%%%%%%%%%%%%%%%%%%%%%%%%%%%%%%%%%%%%%%%%%%%%%%%%%%%%%%%%%
%%%%%%%%%%              BOXPLOTS             %%%%%%%%%%%%%%%%%%%%%%%%%%%%%%%%%%
%%%%%%%%%%%%%%%%%%%%%%%%%%%%%%%%%%%%%%%%%%%%%%%%%%%%%%%%%%%%%%%%

%% If tables they would go here

\end{document}